\documentclass[aps,preprint,floatfix]{revtex4}
\usepackage{amsmath,units,xspace,amssymb,float}
\usepackage[pdftex]{graphicx}
\usepackage{booktabs}

\begin{document}
	

\title{Breakdown of diffusivity$\textrm{-}$entropy scaling in colloidal glass forming liquids}
\author{Bo Li,$^{1,2,3*}$ Xiuming Xiao,$^{1}$ Shuxia Wang,$^{1}$ Weijia Wen,$^{1,2}$ and Ziren Wang$^{1,**}$}
\affiliation{1, Key Laboratory of Soft Condensed Matter Physics and Smart Materials, College of Physics, Chongqing University, Chongqing, 401331, China. \\2, Department of Physics, Hong Kong University of Science and Technology, Clear Water Bay, Hong Kong, China. \\3, Center for Soft and Living Matter, Institute of Basic Science, Ulsan, 44919, South Korea.}
\date{\today}

\begin{abstract} 

Glass is a liquid that has lost its ability to flow. Why this particular substance undergoes its dramatic slowing down in kinetics while remaining barely distinguishable in structure from the fluid state upon cooling constitutes the central question of glass transition physics. Here, we experimentally tested the pathway of kinetic slowing down in glass$\textrm{-}$forming liquids that consisted of ellipsoidal or binary spherical colloids. In contrast to rotational motion, the exponential scaling between diffusion coefficient and excess entropy in translational motion was revealed to break down at startlingly low area fractions ($\phi_\textrm{T}$) due to glassy effects. At $\phi_\textrm{T}$, anormalous translation-rotation coupling was enhanced and the topography of the free energy landscape became rugged. Basing on the positive correlation between $\phi_\textrm{T}$ and fragility, the measurement of $\phi_\textrm{T}$ offers a novel method for predicting liquids' relaxation while circumventing the prohibitive increase in equilibrium times required in high density regions. Our results highlight the role that thermodynamical entropy plays in glass transitions.

\textbf{corresponding emails:} $*$ libotc@gmail.com; $**$ wziren@cqu.edu.cn
\end{abstract}

\maketitle

As a prevalent state of matter, glasses are ubiquitous in nature; various types are manufactured with particular desired mechanical or optical properties on industrial scales, and are widely applied in daily life. However, a consistent microscopic understanding of glassy behaviours remains a challenge for both physicists and material scientists \cite{2011RMP_Berthier}. One central issue in glass science is kinetic slowing down, the process in which a liquid loses its ability to flow upon cooling. Why this particular substance undergoes a dramatic slowing down in kinetics but does not appear to change in structure during the glass transition has been extensively addressed in past decades \cite{1995S_Angell}.

Theories such as random first order transition \cite{2015RMP_Kirkpatrick} and dynamical facilitation \cite{2011PRX_Keys} have achieved great success in the study of glass transitions. For example, the predicted heterogeneous dynamics and structures have been repeatedly confirmed in both experiments \cite{2011PRL_Zhang,2014NC_Zheng,2015PRL_Mishra, 2016NP_Golde} and simulations \cite{2012NP_Kob,2010NM_Tanaka,2006NP_Shintani}. In an alternative approach, the process of kinetic slowing down is directly connected to the thermodynamical quantities such as enthalpy and entropy \cite{1965JCP_Adam}, which are eventually determined by the topography of the free energy landscape (FEL) \cite{1995Science_Stillinger}; though experimental test of FEL in glassy systems have been rare \cite{2013ARCMP_Stillinger}. In presence of abundance of theories, however, puzzles exist against a thorough understanding of glassy dynamics and thus hinder the application in broader realms. One example is the translation-rotation coupling in molecular system \cite{2005PRL_Chong,2005JCP_Moreno}. How the non-central collisions among the non-spherical motifs in microscopic level influence the kinetics of the macroscopic phase composed by the motifs remained poorly understood. Also, clearer link between kinetics and thermodynamics in the glass transition is yet to establish in experiment \cite{2013PRL_Kob}.

As a measure of inherent structures within a basin \cite{1995Science_Stillinger}, excess entropy ($s^{ex}$) has been served as an excellent descriptor of the FEL topography, and therefore has been used to describe the kinetic slowing down of glass$\textrm{-}$forming liquids \cite{2001N_Sastry}. Additionally, the frequency of collisions, which defines the rate at which the system relaxes, is proportional to the number of accessible configurations \cite{1996N_Dzugutov, 1962AJP_Chapman}. This argument directly leads to a scaling law between the long-termed diffusion constant and excess entropy ($D-s^{ex}$ scaling) as $D\sim e^{-\alpha s^{ex}}$. Although the $D-s^{ex}$ scaling for rotational degree of freedom (superscript `$\theta$') has remained unexplored, the $D\sim e^{-\alpha s^{ex}}$ relation in the translational degree of freedom (superscript `$T$' in this paper) has been verified in atomic \cite{2000PRL_Hoyt}, granular \cite{2015PRE_Wang} and colloidal liquids \cite{2001PRL_Samanta,2013PRL_Ma,2015PRL_Thorneywork}. More valuably, this exponential scaling ($D\sim e^{-\alpha s^{ex}}$) \cite{1977PRA_Rosenfeld,1996N_Dzugutov} has been proverbially utilised as a tool to unveil liquid properties such as the anomalous behaviours of water \cite{1999N_Ito,2000N_Scala,2001N_Errington}, atom diffusivity in porous materials \cite{2013Langmuir_Liu}, confinement effect in fluids \cite{2008PRL_Goel}, the relaxation of glass$\textrm{-}$forming liquids \cite{2014PRL_Bnerjee} and quasicrystals \cite{1996N_Dzugutov}; it has become a standard feature of a liquid. However, this $D-s^{ex}$ scaling has only been verified for liquids at temperatures considerably lower than the melting point ($\phi_\textrm{m}$) or glass transition point ($\phi_\textrm{g}$). When approaching $\phi_\textrm{g}$, a glass$\textrm{-}$forming liquid becomes dense and dynamically heterogeneous \cite{2002PRL_Dzugutov}. Theoretically, the $D^{T}-s^{T}$ scaling takes a more precipitous form as $D\sim e^{\frac{\alpha}{Ts^{ex}}}$ near $\phi_\textrm{g}$ \cite{1965JCP_Adam}, where the superscript `$T$' represents the translational degree of freedom. Consequently, it is reasonable to expect the $D-s^{ex}$ scaling to break down at some point upon cooling, at least for glass$\textrm{-}$forming liquids. Since the $D-s^{ex}$ scaling per se is a joint product of thermodynamical entropy and dynamical diffusion coefficient, it can be efficacious for the study of the role that thermodynamics plays in the kinetic slowing down during glass transitions \cite{2001N_Sastry}.

Colloids have served as outstanding model systems for the investigation of glass transitions \cite{2013ARCMP_Lu}. Micro particles dispersed in water undergo Brownian motion, which perfectly simulates the diffusion of atoms or molecules \cite{2016NRM_Li}. The motion of each particle can be recorded and digitalised using optical video microscopy \cite{1996ARPC_Murray}. Hence, kinetic information with single$\textrm{-}$particle resolution can always be obtained on these platforms. The typical relaxation time of a colloidal glass is seconds to hours depending on $\phi$ \cite{2014NC_Zheng}, making the systematic testing of glassy dynamics possible. Recently, numerous theoretical models of glass transitions such as dynamical heterogeneity \cite{2011PRL_Zhang, 2014NC_Zheng, 2015PRL_Mishra}, the point-to-set approach \cite{2015NP_Nagamanasa}, and dynamical facilitation \cite{2014PNAS_Mishra} have been evaluated in colloidal systems.

In this study, we systematically investigated the kinetic slowing down of two crystallising and eight glass$\textrm{-}$forming liquids by measuring their $D-s^{ex}$ scaling (Fig.~1). The colloidal samples consisted of ellipsoidal (with aspect ratio $p=1.26, 1.60, 1.84, 2.68, 3.72, 5.00,$ and $7.06$) or binary spherical particles. More details about the experimental system are presented in the `Methods' section and SI. We found the $D-s^{ex}$ scaling broke down for glass forming liquids (Fig.~2) at unexpectedly low area fractions. In accompany with the breakdown, we observed enhanced translation-rotaion coupling (Fig.~3) and a topography change of the FEL (Fig.~4). Further analysis of relaxation time manifested a positive correlation between the liquids' fragility and the area fraction ($\phi_\textrm{T}$) at which the $D-s^{ex}$ scaling broke down (Fig.~5). Our experiments shed lights onto the studies of molecular glasses and emphasise the crucial roles that thermodynamical entropy play in the kinetic slowing down during the glass transition.

\textbf{Kinetic pathways and $\bf{\textit{D}}~\textrm{-}~\textit{s}^{\textit{ex}}$ scaling}

As the area fraction increases, two kinetic pathways exist for a liquid (Fig.~S2). One way (c-path) is to crystallise into an ordered solid (Fig.~S2b) through a thermodynamic phase transition. When crystallisation of a system is suppressed either through rapid cooling or through incompatible motif symmetry, however the system's glass transition intervenes \cite{1995S_Angell} (Fig.~S2c). The glass transition is abbreviated as `g-path' in this paper. In our experiments, spheres will and ellipsoids with $p=1.26$ were able to follow the c-path; however, binary and other ellipsoids tended to follow the g-path.

In experiments, it is difficult to determine the full value of excess entropy. The two$\textrm{-}$body part of excess entropy ($s^{ex}_\textrm{2}$) was reported to contribute more than 70$\%$ of $s^{ex}$ and have a linear relation with $s^{ex}$, discovered using simulation \cite{2001N_Martinez,2014PRL_Bnerjee}. Hence, we abbreviate $s^{ex}_\textrm{2}$ as $s_\textrm{2}$ in brief as a specific quantity to test the slowing down kinetics. The two-body excess entropies in the translational and rotational degrees of freedom were respectively evaluated as \cite{1989PRA_Baranyai,2014NC_Zheng},
\begin{equation}
s_\textrm{2}^{T}=-\pi k_\textrm{B}\rho \int_\textrm{0}^{+\infty}[g(r)lng(r)-g(r)+1]rdr
\label{01}
\end{equation}

\begin{equation}
s_\textrm{2}^{\theta}=-\dfrac{1}{2}k_\textrm{B}\rho \int_\textrm{0}^{+\infty}g(r)rdr\int_\textrm{0}^{2\pi}g(\theta|r)ln[g(\theta|r)]d\theta
\label{02}
\end{equation}
where $k_\textrm{B}$ is the Boltzmann constant, $\rho$ is the number density, $g(r)$ is the radial distribution, and $g(\theta|r)$ is the orientation distribution function of the angular difference between the long axes at some distance $r$.

In the present study, regarding the rotational degree of freedom, the $D^{\theta}-s_\textrm{2}^{\theta}$ scaling held for all ellipsoids (Fig.~1a) when $\phi$ ranged all the way up close to the orientational glass transition point \cite{2014NC_Zheng}, after which the long time diffusion coefficient vanished. The $D^{\theta}-s_\textrm{2}^{\theta}$ data for rotational motion were adequately fitted by the exponential decay $D^{\theta}\sim e^{-\alpha s_\textrm{2}^{\theta}}$. It's noteworthy that there was a sudden jump of the fitted exponent $\alpha$ between $p=3.27$ and $p=5.00$. Since $\alpha$ reflects how fast the kinetics of the system slows down after losing same number of available configurations, lower $\alpha$ value for small $p$ ellipsoids implies milder and probably continuous increasing of non-ergodicity as a function of $\phi$ \cite{2005PRL_Chong,2005JCP_Moreno}. This observation coincides well with previous molecular-dynamics simulation and thus offers experimental verification of the proposed weak steric hindrance scenario for the slighly elongated particles \cite{2005PRL_Chong,2005JCP_Moreno}.

\begin{figure}[!h]
\centering
\includegraphics[width=1.0\columnwidth]{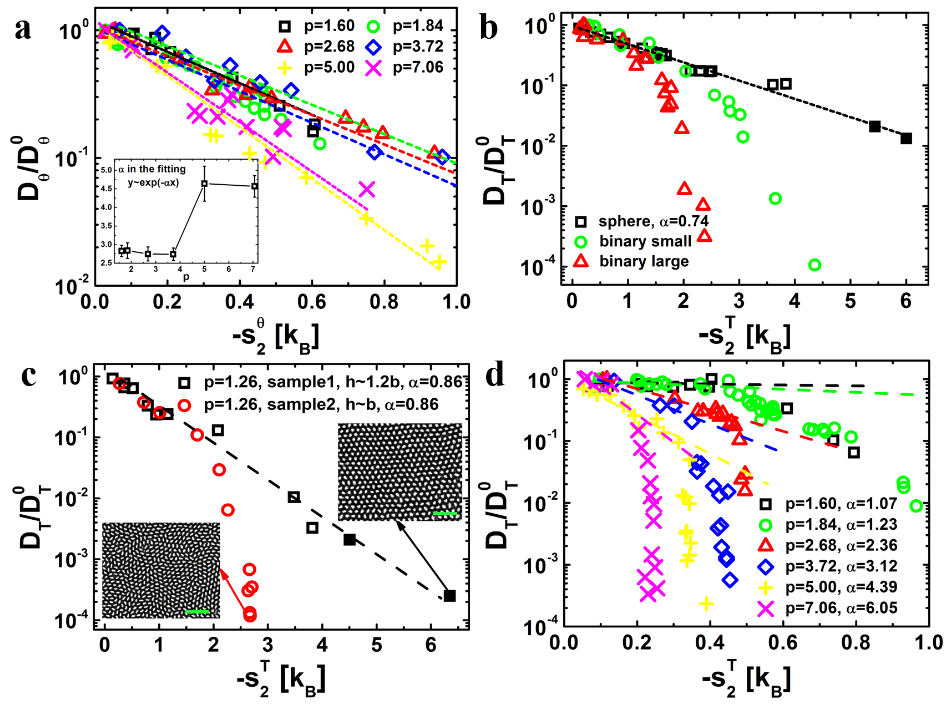}
\caption{\textbf{Relation between diffusion coefficient and two-body excess entropy.} \textbf{a}, $D^{\theta}-s_\textrm{2}^{\theta}$ relation for the rotational motion of ellipsoids with aspect ratios ranging from $1.60$ to $7.06$. The fitted decay exponential is presented in the inset of \textbf{a}. The data for $p=1.26$ ellipsoid is missing because the determination of particles' orientation is inaccurate when they tilted in z-direction. Compared with the case of monodispersed spherical particles in \textbf{b}, the exponential scaling for translational motion ($D^{T}\sim e^{-\alpha s_\textrm{2}^{T}}$) breaks down at certain points for binary spherical (\textbf{b}) and ellipsoidal (\textbf{c, d}) systems. The fitted decay exponential $\alpha$ is presented in the legend of the figures accordingly. In particular, \textbf{c} shows the bifurcation of the $D^{T}-s_\textrm{2}^{T}$ relation for a small$\textrm{-}$aspect$\textrm{-}$ratio ($p=1.26$) ellipsoid system, which either crystalises (right inset of \textbf{c}) or vitrifies (left inset of \textbf{c}), depending on the sample thickness ($h$). The solid points in \textbf{b} and \textbf{c} denote the samples crystalise from their liquid states. The scale bar is 10~$\mu$m. The dashed lines are fitted to the relation $D\sim e^{-\alpha s_\textrm{2}}$.}
\label{fig:1}
\end{figure}

The situation for the translational degree of freedom was qualitatively different. We discovered that for all liquids that follow the c-path (spheres and $p=1.26$ ellipsoids), the $D^{T}-s_\textrm{2}^{T}$ scaling in the translation was valid (Fig.~1b, c) for area fractions ranging from $0.05$ to $0.70$. However, the $D^{T}-s_\textrm{2}^{T}$ scaling broke down for g-path liquids (binary and all ellipsoids) at surprisingly low area fractions ($\phi_\textrm{T}$) (Fig.~2) relative to $\phi_\textrm{g}$ (inset of Fig.~5b). When $\phi$ was higher than $\phi_\textrm{T}$, the kinetics of glass$\textrm{-}$forming liquids slowed down much more rapidly than the kinetics of the liquids that crystalised. Interestingly, for $p=1.26$ ellipsoids, we observed a bifurcated pattern in the $D^{T}-s_\textrm{2}^{T}$ plot (Fig.~1c). Raw images show that the $p=1.26$ ellipsoid crystalised (right inset of Fig.~1c) or formed glass (left inset of Fig.~1c) depending on the wall separation of the samples (Fig.~S2 and related discussions in SI). Consistently, the samples following the c-path obeyed the normal $D^{T}-s_\textrm{2}^{T}$ scaling whereas the samples following the g-path contributed to the data points that deviated from the exponential fitting (dashed lines in Fig.~1c). Basing on the above observations, we conclude that the breakdown of $D^{T}-s_\textrm{2}^{T}$ scaling was caused by the glassy effects.

\begin{figure}[!h]
\centering
\includegraphics[width=1.0\columnwidth]{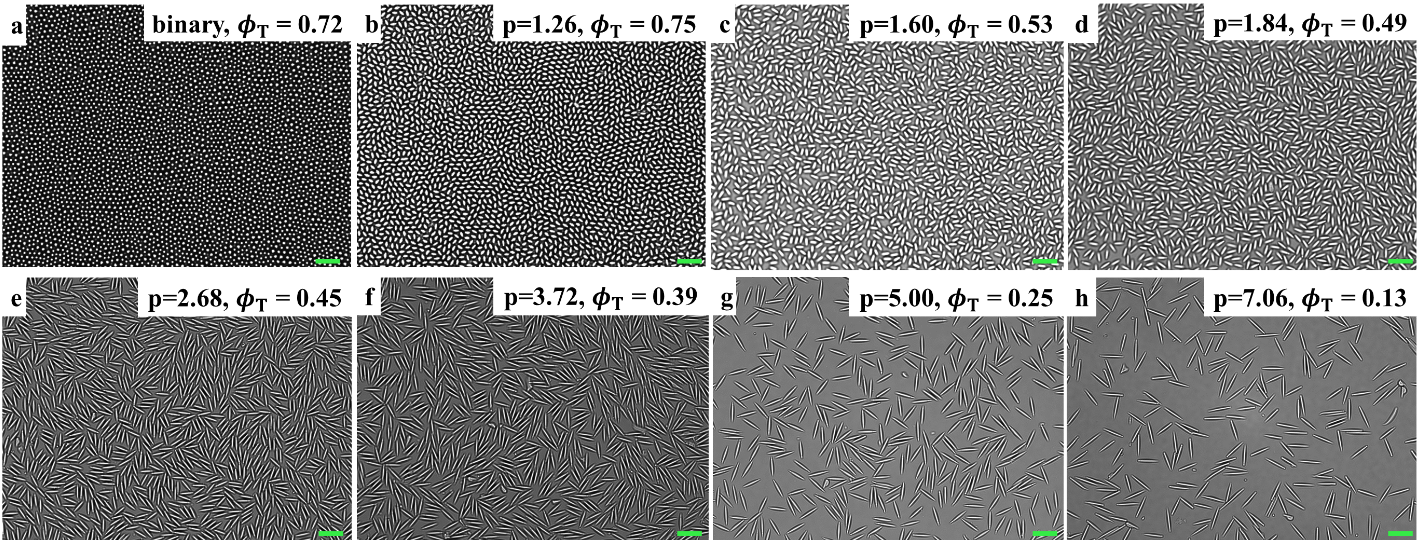}
\caption{\textbf{Raw images captured at the turning points of the translational $\textbf{\textit{D}}^{\textbf{\textit{T}}}-\textbf{\textit{s}}_\textrm{2}^{\textbf{\textit{T}}}$ plots in Fig.~1b-d.} The area fractions ($\phi_{T}$) corresponding to the turning points are specified in each figure. The scale bar is 5~$\mu$m.}
\label{fig:2}
\end{figure}

There existed a turning point, which was defined as the first point in the $D^{T}-s_\textrm{2}^{T}$ plots (Fig.~1b$\textrm{-}$d) that deviated from the exponential fitting (dashed line in Fig.~1b$\textrm{-}$d); this point thus separated the smooth and rapid slowing down processes in the translational degree of freedom for each g-path liquid. The raw images corresponding to the turning points and the area fractions ($\phi_\textrm{T}$) are presented in Fig.~2. Confusingly but also interestingly, $\phi_\textrm{T}$ covered a wide $\phi$ range from $0.13$ to $0.72$ for different g-path liquids. For the $p=1.26$ ellipsoid (Fig.~2b), $\phi_\textrm{T}$ was higher than the area fraction at which dynamical heterogeneity (DH) occurs \cite{2015PRL_Mishra} (Figs.~2c, S3, S6b, S7b), whereas for binary (Figs.~S2b, S4, S6a, S7a) and other ellipsoids \cite{2015PRL_Mishra,2014NC_Zheng} (Fig.~2c-h), $\phi_\textrm{T}$ was (much) lower than that at the onset of DH (Figs.~S4-S7). Therefore, any turning in the $D^{T}-s_\textrm{2}^{T}$ scaling should have been irrelevant to the DH. The properties (such as $\phi_\textrm{T}$) of transient region between the liquid and glass states are usually overlooked by current dynamical theories, in part because the density of a system in this region is apparently too low to allow any DH to occur. Instead of DH, three major features of the system at $\phi_\textrm{T}$ will be reported in the following sections.  

\textbf{Abnormal translational$\textrm{-}$rotational cross$\textrm{-}$correlation at $\boldsymbol{\phi}_\textrm{T}$}

We discovered that the abnormal coupling between translation and rotation was significantly enhanced at $\phi_\textrm{T}$. For systems composed of anisotropic shaped particles such as ellipsoids \cite{2014NC_Zheng} or dumbbells \cite{2010JCP_Chopra}, the translational and rotational motions are no longer independent whenever a collision occurs. Such systems are usually also effective glass formers. However, the relationship between $\phi$ and collision frequency, which can be calculated from the height of the first peak of the radial distribution function $g(r)$ \cite{1962AJP_Chapman, 1996N_Dzugutov}, can be complicated if a system takes its g-path. For a strong glass former in particular, the shape of $g(r)$ hardly changes with increasing $\phi$ (Fig.~S15c) \cite{1983PRB_Steinhardt}. Therefore, how the enhanced translational$\textrm{-}$rotational correlation affects the kinetic pathway and eventually alters the $D^{T}-s_\textrm{2}^{T}$ scaling remains unexplored.

\begin{figure}[!h]
	\centering
	\includegraphics[width=1.0\columnwidth]{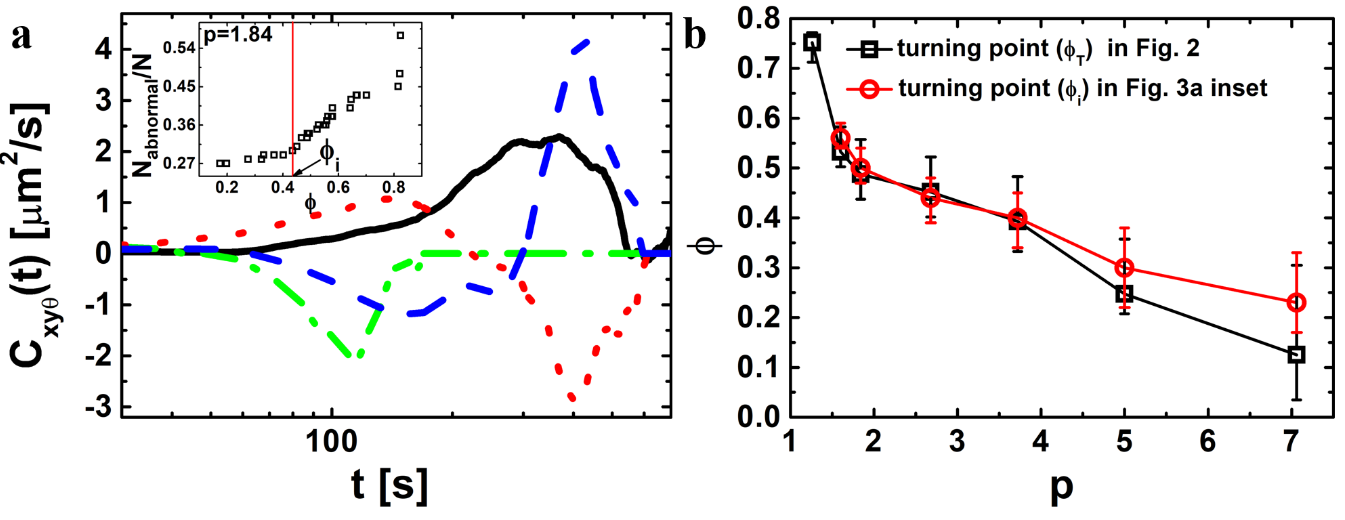}
	\caption{Abnormal cross$\textrm{-}$correlation in ellipsoid systems. \textbf{a,} Four types of cross$\textrm{-}$correlation $C_{xy\theta}(t)$ in ellipsoid systems. The fraction of abnormally correlated (red-dot, blue-dash, and green-dash-dot lines) particles as a function of area fraction \textbf{(inset)} starts to increase at $\phi_\textrm{i}$. \textbf{b,} Area fraction corresponding to the turning point ($\phi_\textrm{T}$) in $D^{T}-s_\textrm{2}^{T}$ (\textbf{Fig.~1c,d}) and abnormal correlation ($\phi_\textrm{i}$ in \textbf{a (inset)}).}
	\label{fig:3}
\end{figure}

To characterise the translational$\textrm{-}$rotational coupling, we defined the cross$\textrm{-}$correlation function for ellipsoidal particles as $C_{xy\theta}(t)=2\langle\Delta x\Delta y\rangle \langle sin2\theta \rangle /t$, where $\Delta x$ and $\Delta y$ are displacements in time period $t$ in $x$ and $y$ directions, respectively; $\theta$ is the angle of an ellipsoid (in lab frame) at time $t$; the angle brackets represent the ensemble average. For a single ellipsoid diffusing in water, the friction coefficient parallel to the major axis is invariably smaller than that perpendicular to the major axis, resulting in positive translation$\textrm{-}$rotation coupling in the lab frame \cite{2006S_Han}. If an ellipsoid particle is in a crowded environment such as a suspension of high density active bacteria, however, negative cross correlation may be observed \cite{2016PRL_Peng}. In our experiments, we observed four types of cross$\textrm{-}$correlation in total (Fig.~3a). Aside from the normal case for a single Brownian ellipsoid (black curve in Fig.~3a), the other three types were abnormal because the correlation was negative during certain time periods.

We counted the number fraction of particles with abnormal cross$\textrm{-}$correlation for each $\phi$ and for all seven types of ellipsoids; we found that the proportion of abnormal particles started to mushroom at certain area fractions $\phi_\textrm{i}$ (inset of Fig.~3a). In addition, $\phi_\textrm{i}$ coincided favorably with $\phi_\textrm{T}$ within the experimental accuracy (Fig.~3b). As previously discussed, the negative $C_{xy\theta}(t)$ was caused by abnormally slow diffusion along the major axis, which is strictly prohibited for a single Brownian ellipsoid. As $\phi$ increased, the abnormal diffusion could thus only have been generated by increasingly frequent collisions among particles. Thus, the booming ratio of particles with abnormal cross-correlation at $\phi_\textrm{i}$ was an effective indicator of enhanced collision frequency within the system. In other words, $\phi_\textrm{i}$ is also the point at which particles started to collide frequently. Hence, it seems that the observed abnormal cross$\textrm{-}$correlation strongly indicated that the microscopic origin causing the breakdown of $D^{T}-s_\textrm{2}^{T}$ scaling in ellipsoidal systems could be deeply rooted in the translation-rotaion coupling caused by noncentral collisions among particles.

Nonetheless, if the turning in $D^{T}-s_\textrm{2}^{T}$ scaling was solely caused by translation$\textrm{-}$rotation coupling, it should have commonly existed for both degrees of freedom, which is contrary to what was observed in the experiments (Fig.~1a). The absence of turning in the rotational degree of freedom, as well as the existence of turning in binary systems creates ambiguity regarding the role played by collision in the turning. In addition, it is questionable whether the short-timescaled ($\sim 100~s$ in Fig.~3a) cross-correlation (or collisions) would be able to influence the relaxation of glasses at much longer timescales ($>1000s$). Therefore, the roles that noncentral collisions and the translation$\textrm{-}$rotation coupling play in the kinetic slowing down of ellipsoidal systems remain an open question that warrants further investigation.

\textbf{Rugged FEL topography near $\boldsymbol{\phi}_\textrm{T}$}

Another feature of the turning is the rugged FEL topography near $\phi_\textrm{T}$. For a system with $N$ particles, the FEL is a $6^{N}$ hypersurface, which is hardly calculable when $N$ is large \cite{1997PRL_Heuer}. In practice, it was proven by simulation that average particle mobility can approximately depict the topography of any FEL \cite{2000PRL_Buchner,2006PRL_Appignanesi,2003PRL_Denny}. Herein, we express the mobility in translational motion as $\Delta u(t)=(\frac{1}{N})\sum_\textrm{i=1}^{N}[\textbf{u}_\textrm{\textbf{i}}(t+\tau_{\alpha}/2)-\textbf{u}_\textrm{\textbf{i}}(t-\tau_{\alpha}/2)]$, where $N$ is the particle number, $\tau_{\alpha}$ is the relaxation time (defined in Fig.~5a), and $\textbf{u}_\textrm{\textbf{i}}$ is particle $i$'s displacement at a given time \cite{2000PRL_Buchner,2006PRL_Appignanesi,2003PRL_Denny}. Simulations suggested that over a much broader $\phi$ range than the range in which mode coupling theory takes effect, the relaxation kinetics are sensitive mainly to the topography of the FEL \cite{1997PRL_Heuer, 2006PRL_Appignanesi, 2003PRL_Denny, 2003PRL_Doliwa}, implying that inherent structure is a useful tool with which to predict glassy kinetics when the density of the system is low. Fig.~4a-c we display the time dependence of $\Delta u(t)$ for the $p=5.00$ ellipsoid under three typical area fractions. Compared with the FEL at low density (Fig.~4a), the topography of the FEL in translational motion became rugged in the vicinity of $\phi_\textrm{T}$ (Fig.~4b) and thereafter (Fig.~4c). In contrast, the topography of the FEL in rotational motion, $\Delta\theta (t)=(\frac{1}{N})\sum_\textrm{i=1}^{N}[\theta_\textrm{\textbf{i}}(t+\tau_{\alpha}/2)-\theta_\textrm{\textbf{i}}(t-\tau_{\alpha}/2)]$, remained smooth over the whole $\phi$ range (Fig.~4d-f), in agreement with the absence of turning point in the $D^{\theta}-s_\textrm{2}^{\theta}$ plot (Fig.~1a). These observations confirmed the conclusive role the inherent structures play in $D-s_\textrm{2}$ scaling. The evolution of the FEL topographies of all other g-path liquids exhibited tendencies (Figs.~S9-S14) similar to that shown in Fig.~4. To completely confirm the causal role FEL plays in the breakdown of the $D^{T}-s_\textrm{2}^{T}$ scaling, however, required a more direct access to inherent structures. 

\begin{figure}[!h]
\centering
\includegraphics[width=1.0\columnwidth]{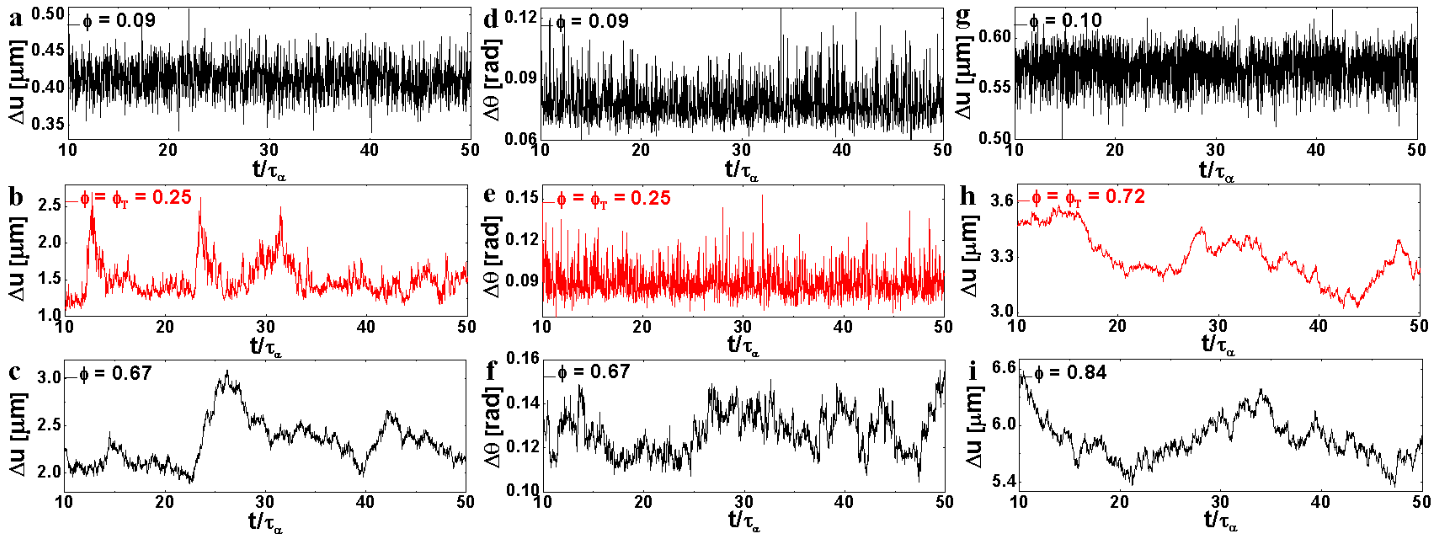}
\caption{\textbf{Topography of the FEL in representation of particle mobility.} For the $p=5.00$ ellipsoid system, we chose \textbf{a}, $\phi=0.09$; \textbf{b}, $\phi=\phi_\textrm{T}=0.25$ and \textbf{c}, $\phi=0.67$ samples and calculated the time evolution of their translational displacements, $\Delta u(t)$, during the relaxation time $\tau_{\alpha}$. The rotational displacements $\Delta\theta (t)$ under same $\phi$ are presented in \textbf{d-f}. The translational displacements of binary samples with \textbf{g}, $\phi=0.10$; \textbf{h}, $\phi=\phi_\textrm{T}=0.72$ and \textbf{i}, $\phi=0.84$ are also presented. $\tau_{\alpha}$ is defined as the time at which the intermediate scattering function in Fig.~5a has decayed to $1/e$.}
\label{fig:4}
\end{figure}

Near $\phi_\textrm{T}$, some adjacent inherent structures formed basins on larger time scales (Fig.~4b, c), during which the system was trapped within a small number of inherent structures. The presence of this substructure in that FEL reflected the highly nontrivial nature of the entire hypersurface as resulting from many-body interactions \cite{1984S_Stillinger}. When $\phi$ was high, the presence of valleys dramatically slowed down the kinetics and was hence the limiting step for full structural relaxation. During residence in a basin, the configuration fluctuated only slightly and thus corresponded to a long-lived metastable state, which abrogated the ergodicity of the system. High activation energy was required for the system to relax from one basin to another \cite{2001N_Debenedetti}. In other words, fewer and fewer configurations were available after $\phi_\textrm{T}$. As a consequence, when $\phi$ increases, the loss of configurations (or entropy, $s_\textrm{2}^{T}$) leads to much more violent slowing down in kinetics ($D^{T}$) compared with the situation before $\phi_\textrm{T}$. When $\phi$ was low, however, the residence probability in the configuration space was no longer dominated by the regions around the inherent structures; therefore no stationary inhomogeneous density field could be defined over periods longer than microscopic times \cite{2001N_Debenedetti}. Thus the kinetics became irrelevant to the topography of the FEL \cite{1995Science_Stillinger}.

\textbf{Positive correlation between fragility and $\boldsymbol{\phi}_\textrm{T}$}

A more crucial feature of $\phi_\textrm{T}$ is its connection with the fragility ($m$), another quantity that reflects the inherent character of the FEL \cite{1995S_Angell}. To determine the fragility of a liquid, we performed the following steps successively. First of all, we calculated the intermediate scattering function $F_\textrm{s}(\textbf{q},t)=\langle \sum_\textrm{j=1}^{N} e^{i\textbf{q}\cdot(\textbf{x}_\textrm{j}(t)-\textbf{x}_\textrm{j}(0))}\rangle /N$ (Fig.~5a). Here, $\textbf{x}_\textrm{j}(t)$ is the position of ellipsoid $j$ at time $t$, $\textbf{q}$ is the scattering vector, $N$ is the total number of particles, and $\langle \rangle$ denotes the ensemble average. We chose $q=q_{m}$ corresponding to the first peak in the structure factor at high density. The relaxation time ($\tau_{\alpha}$) was then defined as the time at which $F_\textrm{s}(\textbf{q},t)$ had decayed to $1/e$ (dotted line in Fig.~5a). Secondly, we plotted $1/\tau_{\alpha}$ as a function of $\phi$ and then extrapolated $1/\tau_{\alpha}$ to zero (Fig.~5b) with the equation $1/\tau_{\alpha}=(\phi_\textrm{g}-\phi)^{\gamma}$. The relaxation time, according to this formula, must diverge at $\phi_\textrm{g}$, which serves as the glass transition point of the system. Our fitted $\phi_\textrm{g}$ (Fig.~5b inset) decreased with aspect ratio, and our values were consistent with the those reported from previous colloid experiments \cite{2011PRL_Zheng, 2014NC_Zheng}. Finally we produced Arrhenius plots of relaxation time as a function of area fraction scaled by values of $\phi_\textrm{g}$ (Fig.~5c). The fragility $m$ was then defined as the derivative of the scaled relaxation time as a function of area fraction scaled by $\phi_\textrm{g}$, i.e. $m=\frac{\partial(\tau_\alpha/\tau_{\alpha}^{0})}{\partial(\phi/\phi_\textrm{g})}\mid_{\phi=\phi_\textrm{g}}$  \cite{2012RPP_Hunter}.

\begin{figure}[!h]
\centering
\includegraphics[width=1.0\columnwidth]{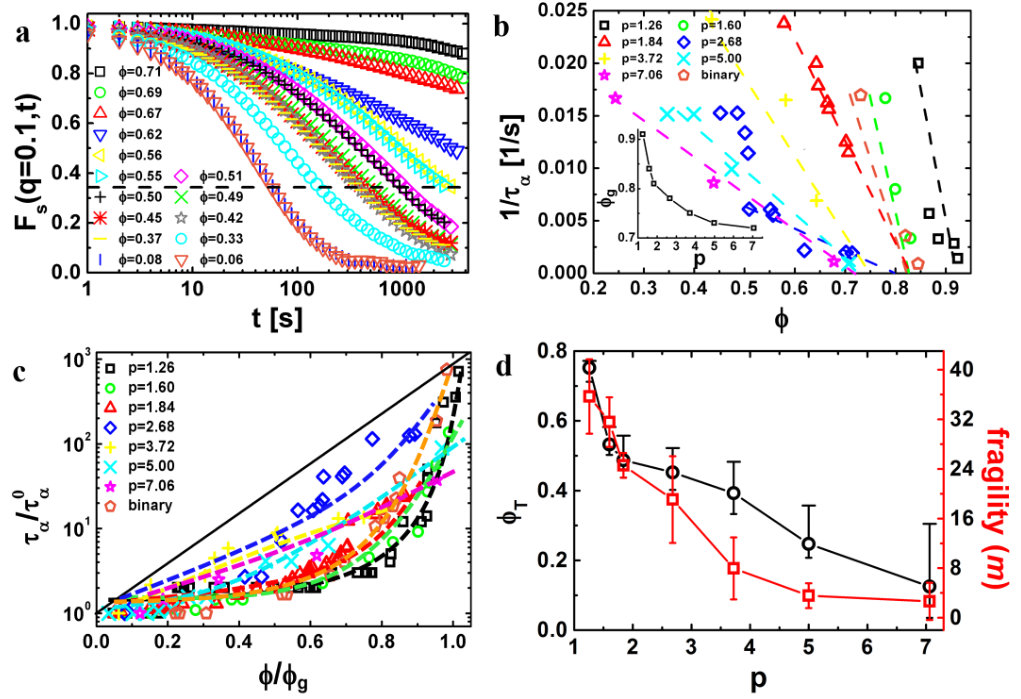}
\caption{\textbf{Fragility and $\textit{\textbf{D}}^{\textbf{\textit{T}}}-\textit{\textbf{s}}_\textrm{2}^\textit{T}$ turning.} \textbf{a}, The intermediate scattering function, $F_\textrm{s}(\textbf{q},t)$, of the ellipsoid with $p=2.68$. The relaxation time was defined as the time at which $F_\textrm{s}(\textbf{q},t)$ decayed to the value $1/e$ (dashed horizontal line). \textbf{b,} The inverse of the relaxation time as a function of area fraction for all glass$\textrm{-}$forming liquids composed of ellipsoid particles. The glass transition points ($\phi_\textrm{g}$) were determined by extrapolating (dashed lines) the inversed relaxation time to zero. $\phi_\textrm{g}$ of ellipsoids as a function of $p$ are shown in the inset of \textbf{b}. \textbf{c}, Arrhenius plots of relaxation time as functions of area fraction scaled by $\phi_\textrm{g}$. The dashed lines are for better guidance of looking. \textbf{d}, Both the fragility ($m$) and turning point ($\phi_\textrm{T}$) of the $D-s_\textrm{2}$ relation decreased with aspect ratio.}
\label{fig:5}
\end{figure}

As is clear in Fig.~5d, both $m$ and $\phi_\textrm{T}$ decreased with aspect ratio. Without exception, the data of the binary system conformed to this tendency. Hence the striking differences between fragile and strong liquids can be directly evaluated using the $D^{T}-s_\textrm{2}^{T}$ scaling (Fig.~1b-d). The kinetic slowing down of a strong liquid begins to deviate from the form $D\sim e^{-\alpha s^{ex}}$ at a much lower $\phi_\textrm{T}$ than does a fragile liquid. A molecular dynamics (MD) simulation showed that, when $\phi$ increasing, $D$ decayed faster for strong liquids than it did for fragile liquids \cite{2001N_Voivod}. Our observations provided clear experimental support to this finding (Fig.~1d). The difference between strong and fragile liquids can also be evaluated from the structural reaction of a liquid towards cooling. Usually, a stronger liquid exhibits less structural change within the same temperature range \cite{2014NC_Mauro}. Fig.~S15 illustrates that for a similar range of diffusion coefficients, decreasing from $1$ to $10^{-3}$, the first peak of the radial distribution function ($g(r)$) of fragile liquids altered correspondingly (Fig.~S15a, b, d) whereas it hardly changed for strong liquids (Fig.~S15c, d). There also existed evidence that the first peak of $g(r)$ saturated at $\phi_\textrm{T}$.

The unambiguous positive correlation between $\phi_\textrm{T}$ and $m$ (Fig.~5d) highlights the crucial role of $\phi_\textrm{T}$ in glass transitions. For a g-path liquid, there exists a continuous transformation from fast relaxation in the low $\phi$ region to slow relaxation in the high $\phi$ region (Fig.~5c). The transformation usually occurs at lower $\phi$ regions for strong liquids than it does for fragile liquids. This explains why $\phi_\textrm{T}$ was lower for ellipsoids with larger p, or equivalently, stronger liquids in our experiments (Figs.~2b-d, 5d). In addition, since a liquid's relaxation is characterised by $m$ under the thermodynamic picture, the relaxation of a g-path liquid at high $\phi$ can be precisely predicted by measuring its $D^{T}-s_\textrm{2}^{T}$ relation and identifying $\phi_\textrm{T}$. Not like $\phi_\textrm{g}$, $\phi_\textrm{T}$ usually lies well within liquid region. Therefore, $\phi_\textrm{T}$ should be useful for probing glassy behaviours that are typical of much higher $\phi$ while circumventing the prohibitive increase in equilibration times.

The slowing down kinetics after $\phi_\textrm{T}$ were also directly related to $m$ and could be understood from the visualizations of FEL \cite{1995Science_Stillinger}, in which the variation in behaviours between the strong and fragile liquids was tracked back to topographic differences in the FEL. The extreme of strong glass formers ($p=3.72, 5.00$, and $7.06$ ellipsoids in our experiments, Fig.~5c) presented a uniformly rough topology, with little or no coherent organisation of the individual basins into large and deep craters \cite{1995Science_Stillinger}. The relaxation of the system was dominated by jumps among a few isolated local minima (basin of attraction) in a single energy$\textrm{-}$scaled FEL \cite{2000PRL_Buchner, 1997PRL_Heuer}. In other words, the configurations in which the system could dwell were dearth for a strong liquid. As the area fraction increased, the local minima became deeper and the relaxation between any two minima required overcoming an increasingly high energy barrier. The system remained within one local minimum. Therefore, the loss of local minima (or entropy) upon cooling led to much faster slowing down in the kinetics of a strong liquid (Fig.~5c), as is reflected by the steeper $D^{T}-s_\textrm{2}^{T}$ scaling after $\phi_\textrm{T}$ in Fig.~1d.

\textbf{Discussion}

To sum up, we have reported the first systematic experimental study of kinetic slowing down of colloidal glass$\textrm{-}$forming liquids from the perspective of thermodynamics \cite{1985JCP_Stillinger, 1999PRL_Sciortino}. We discovered the $D^{T}-s_\textrm{2}^{T}$ scaling for all g-path liquids turned at fairly low area fractions ($\phi_\textrm{T}$), which defined a new `transition' point that has not been addressed in any previously published theory. After $\phi_\textrm{T}$, the slowing down process was greatly accelerated. This observation demonstrates the crucial role thermodynamic entropy plays in kinetic slowing down. In addition, there existed profound interactions among kinetic slowing down, translation-rotation coupling, the FEL topography and fragility.

The most challenging task accompanied with this research is how to include our observations with current theoretical frameworks \cite{2011RMP_Berthier}. Although not in conflict with any well-established model, most of our observations have not been addressed by established theories at all. The turning point ($\phi_\textrm{T}$) proposed in this article has been proven to be irrelevant with the onset of DH (Figs.~2, S3-S7) and was too low to attract any attention in previous theories \cite{2011RMP_Berthier}. Meanwhile, our findings strongly indicate that behaviors of low density liquids can foreshadow the kinetic paths and relaxation phenomena of their high density counterparts, namely, glasses. It has been reported that some rugged FEL occurred when $T$ was still much higher than the transition point predicted by the mode coupling theory \cite{2003PRL_Doliwa}. Our observations that both rugged FEL and $D^{T}-s_\textrm{2}^{T}$ turning (i.e. $\phi_\textrm{T}$) occurred much earlier than DH appear to support this viewpoint. Furthermore, we proved that the origin of glassy behaviours was deeply rooted in the properties of the low density liquids that originally populated the system, far before the glass transition. Since $\phi_\textrm{T}$ is essentially a `blended' quantity that determined by the joint effect of entropy and diffusion coefficient, it is worthwhile to seek for interplays between thermodynamics and kinetics in theory. 

The reason why the FEL topographies in different degrees of freedom are so distinct was yet to be fully explored. Our data (Figs.~4a-f, S10-S14) unambiguously demonstrated that the FEL in translational motions became rugged in advance of rotational motions, suggesting the ellipsoids were still free to rotate when they were already non-ergodic in translation. This caging sequence support the previous results by MD simulation \cite{2005PRL_Chong, 2005JCP_Moreno} where two glass transition lines were discovered in the dynamical phase diagram. In addition, the jump in the decay exponent of $D^{\theta}-s_\textrm{2}^{\theta}$ in Fig.~1a(inset), as well as the sharp decreasing of $\phi_\textrm{T}$ (Fig.~5d) as a function of $p$ indicate qualitative distinct glass transitions for small and large $p$ ellipsoids due to different translation-rotation couplings. It's therefore meaningful to take more quantitative comparison of experiment and simulation data and investigate the characteristics of strong and weak steric hindrance scenarios. Our current experiments pave the way of establishing a link between random first order transition theory and the inherent structure. 

\textbf{Methods}

In this experiment, we used two types of colloidal glass$\textrm{-}$forming liquids consisting of ellipsoidal or binary spherical particles. For the ellipsoidal system, we stretched polymethyl methacrylate (PMMA) spheres (Microparticles GmbH, Germany) with a diameter of $\sigma=(2.74 \pm 0.04)~\mu$m  into ellipsoid particles with aspect ratios $p=a/b=1.26, 1.60, 1.84, 2.68, 3.72, 5.00$, and $7.06$ and polydispersity of less than 5$\%$. Here, a and b are the major and minor axes, respectively. For binary spherical system, we mixed $\sigma=(2.08 \pm 0.05)~\mu$m and 2.74$~\mu$m PMMA spheres homogeneously with a ratio of $0.55:0.45$. Both the ellipsoidal and binary spherical systems were able to avoid crystallisation effectively and hence served as good glass formers. The ellipsoids with $p=1.26$ sometimes formed rotator crystals \cite{2014NC_Zheng} when the wall separation was appropriate; they formed glass in most trails. We also used 2.74$~\mu$m PMMA mono-dispersed spheres as a control group. To optimally image and track particle motion, we adopted a quasi$\textrm{-}$two$\textrm{-}$dimensional sample configuration, in which colloidal particles were dispersed in water between two parallel horizontal glass walls, the separation between which was less than 1.2~$\sigma$, screening off the majority of the hydrodynamic force between particles \cite{2013PRL_Ma}. For colloids in 2D, the area fraction $\phi=\pi ab\rho$ played the same role as the inverse temperature $1/T$ plays in molecular systems, with $\rho$ the number density. Approximately 10 to 25 area fractions within $0.02\leq\phi\leq 0.95$ were recorded for each aspect ratio through video microscopy. During the 2- to 4-$hours$ measurements at each $\phi$, no drift, flow, or density change were observed. The center-of-mass positions and orientations of $200-5000$ ellipsoids were tracked using an image-processing algorithm \cite{1996ARPC_Murray}.

\textbf{Acknowledgements} The authors sincerely thank Prof. Walter Kob for critical reading of the manuscript and insightful comments. This study was supported by the National Natural Science Foundation of China (Grant No.~11604031) and the Fundamental Research Funds for the Central Universities of China (Grant No. 106112015CDJZR308801).

\textbf{Supplementary Information is at the end of this manuscript}

\newpage

\section{Preparation of samples}

In this work ellipsoidal particles were fabricated using the method described in Ref.~\cite{1993CPS_Ho}. We placed $0.5\%$ PMMA (polymethyl methacrylate) spheres into a $12\%$ aqueous PVA (polyvinyl alcohol) solution in a Petri dish. After the water in the PVA solution had evaporated, we stretched the PVA film at approximately $130^\circ$C. PMMA spheres embedded in the film can be simultaneously stretched because their glass transition temperatures are below $130^\circ$C. After the product had cooled to room temperature, we dissolved the PVA and obtained ellipsoids with $5\%$ polydispersity. The clean ellipsoid solution was stabilised using 7$millimole$ SDS (sodium dodecyl sulfate), after which the samples were sonicated and centrifuged.

\renewcommand{\thefigure}{S1}
\begin{figure}[!h]
	\centering
	\includegraphics[width=0.8\columnwidth]{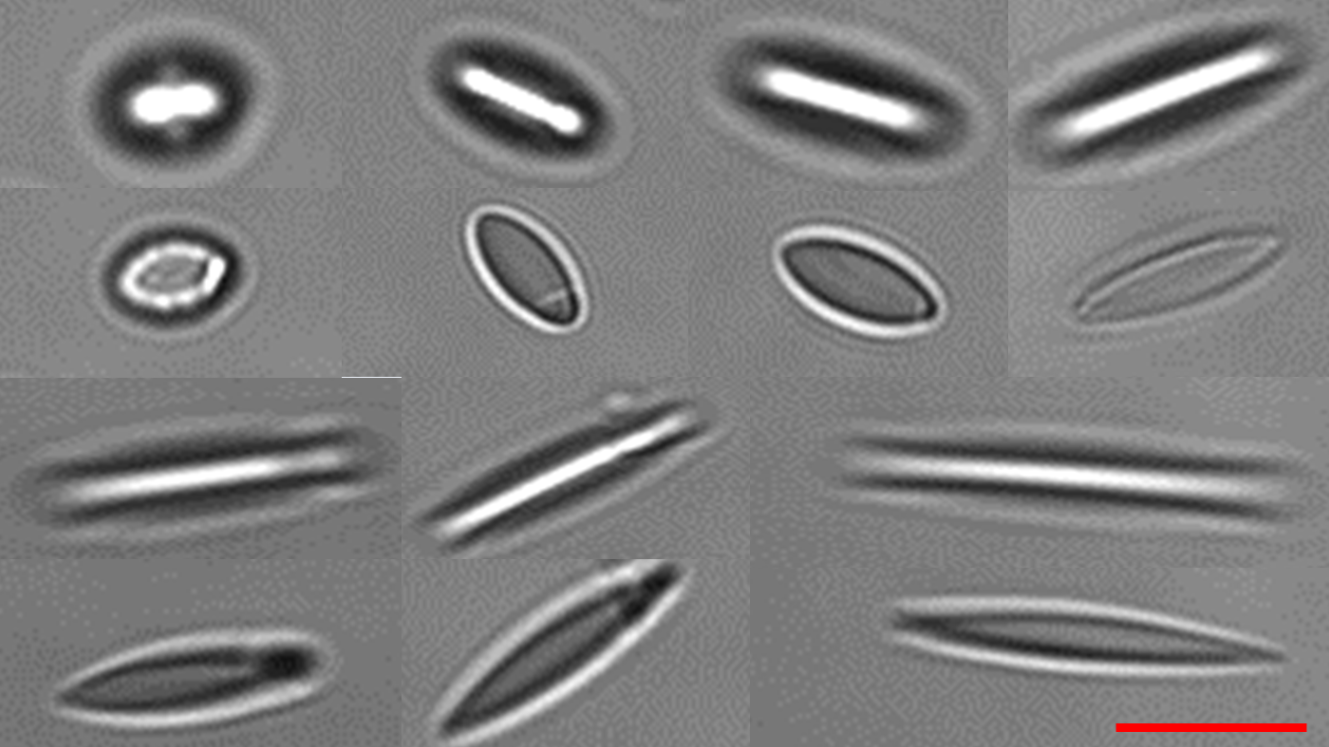}
	\caption{Contrast images of ellipsoids for particle tracking (upper panel) and images that reflect the true aspect ratios of the particles (lower panel). The scale bar is 3~$\mu$m.}
	\label{fig:S1}
\end{figure}

The aspect ratios of the ellipsoids were calculated as follows. If the length of the PVA film was elongated by a factor of $q$, the long axis of the ellipsoid became $a=q\sigma$, where $\sigma$ is the diameter of the spherical particles. The short axis $b$ of the ellipsoid was then calculated using the fact that the volume of a particle was conserved before and after the stretching; that is, $\frac{4}{3}\pi (\sigma/2)^{3}=\frac{4}{3}\pi (a/2)(b/2)^2$. The aspect ratio of the ellipsoids was then $p=a/b$.

The images of ellipsoids that are optimal for particle tracking (upper panel of Fig.~S1) differ from the images that accurately show the aspect ratios of the particles (lower panel of Fig.~S1). In our experiments, we adopted the images in the upper panel of Fig.~S1 for the purpose of image processing. Therefore, the ellipsoids in the raw images appear to have higher aspect ratios than their true values. The choice does not affect the precision of the mass-to-center positions and the angles calculated by the imaging tracking algorithm widely adopted in studies \cite{2006S_Han,2011PRL_Zheng,2014PNAS_Mishra}.

\renewcommand{\thefigure}{S2}
\begin{figure}[!h]
	\centering
	\includegraphics[width=1.0\columnwidth]{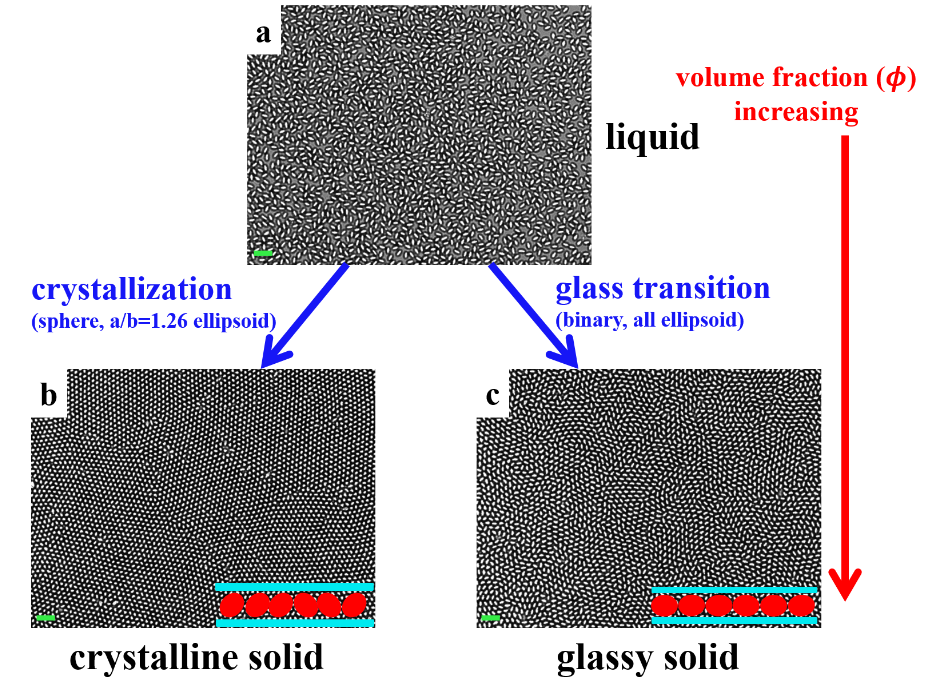}
	\caption{Kinetic pathways of liquids upon cooling. As the volume fraction increases, a liquid either crystalises into \textbf{b} or transforms into a amorphous state in \textbf{c} via a glass transition. The scale bar is 10~$\mu$m.}
	\label{fig:S2}
\end{figure}

For binary spherical systems, we mixed 2.08~$\mu m$ and 2.74~$\mu m$ PMMA spheres with number ratio of $0.55:0.45$. The ratio was chosen based on the standard Kob$\textrm{-}$Anderson model \cite{1993PRE_Kob} and was proverbially chosen in colloidal experiments \cite{2015NP_Nagamanasa}. To ensure the homogeneous mixture of two types of spheres, we shook the sample for 1 min and horizontally settled it for 4 hours on the objective stage before recording data.

Because the diffusion constants at low $\phi$ were sensitive to the separation between the glass walls, we recorded data for all area fractions within the same sample. To achieve this, we tilted the monolayers by a small angle of $0.5^\circ$. After equilibration for weeks, the sedimentation$\textrm{-}$diffusion equilibrium was well established \cite{2017PRL_Thorneywork}. Particle density was homogenous within the field of view and no drift or flow was observed during the experiment period.

As discussed in the main text, a $p=1.26$ ellipsoid (Fig.~1c) was able to take either the c-path (Fig.~S2a, b) or g-path (Fig.~S2a, c), depending on the separation of the two glass walls. For monolayer colloidal samples, the wall separation, $h$, usually satisfied the condition $b<h<1.2b$, where $b$ denotes the short axis of particles in the ellipsoidal systems. When $h$ was close to $b$, the ellipsoids were strictly confined in the xy-plane (inset of Fig.~S2c). As $\phi$ increased, the incompatible motif ruined crystal symmetries. The formation of liquid crystal phases was also avoided due to higher charge density at the steeple of the particles \cite{2014NC_Zheng}. When $h$ approached to $1.2b$, however, the system had free space in which the ellipsoids tilted slightly in the z-direction (inset of Fig.~S2b). Consequently, under high $\phi$, the additional rotational entropy in the z-direction tended to drive the system into the rotator phase \cite{2009JMC_Hosein}, as illustrated in Fig.~S2b. Note that the $p=1.26$ ellipsoids in Fig.~S2b appear smaller and more like spherical particles compared with those shown in Fig.~S2c because what we observed is illustrated as a xy-cross-section of tilted ellipsoids in Fig.~S2b. This observation clearly demonstrates that the ellipsoids were indeed tilted in the z-direction when both $h$ and $\phi$ were relatively high.

\section{Dynamical heterogeneity and $\boldsymbol{\phi}_\textrm{T}$}

\renewcommand{\thefigure}{S3}
\begin{figure}[!h]
	\centering
	\includegraphics[width=1.0\columnwidth]{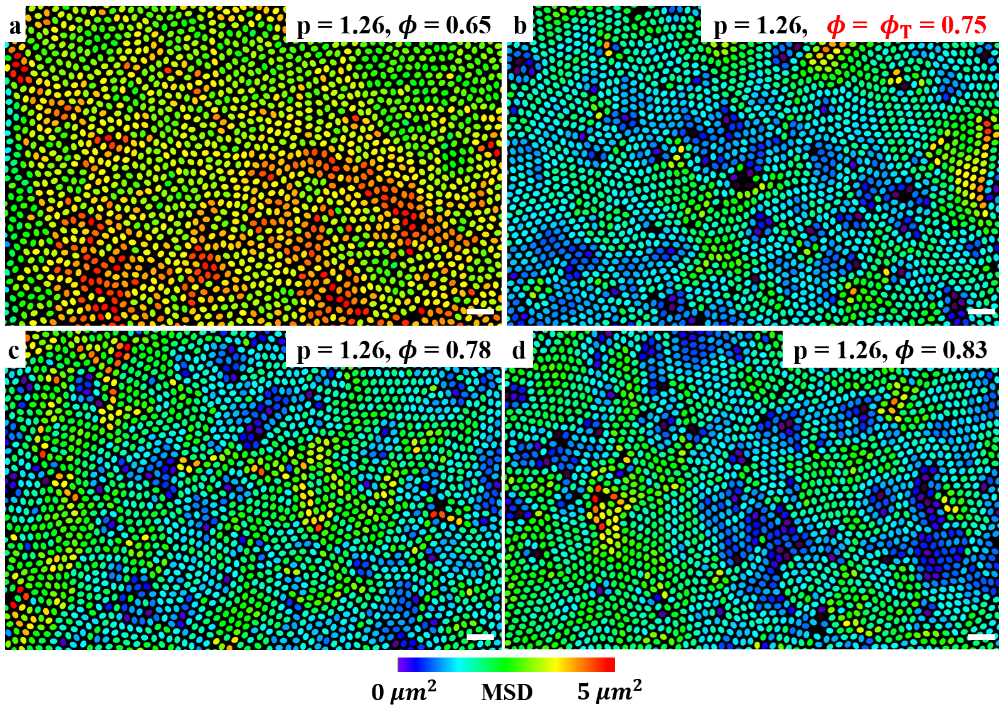}
	\caption{Dynamical heterogenity of $p=1.26$ ellipsoids under \textbf{a} $\phi=0.65$, \textbf{b} $\phi=\phi_\textrm{T}=0.75$, \textbf{c} $\phi=0.78$ and \textbf{d} $\phi=0.83$. The color represents the value of the mean square displacement of a particle during 500~$s$. The scale bar is 10~$\mu$m.}	
	\label{fig:S3}
\end{figure}

\renewcommand{\thefigure}{S4}
\begin{figure}[!h]
	\centering
	\includegraphics[width=1.0\columnwidth]{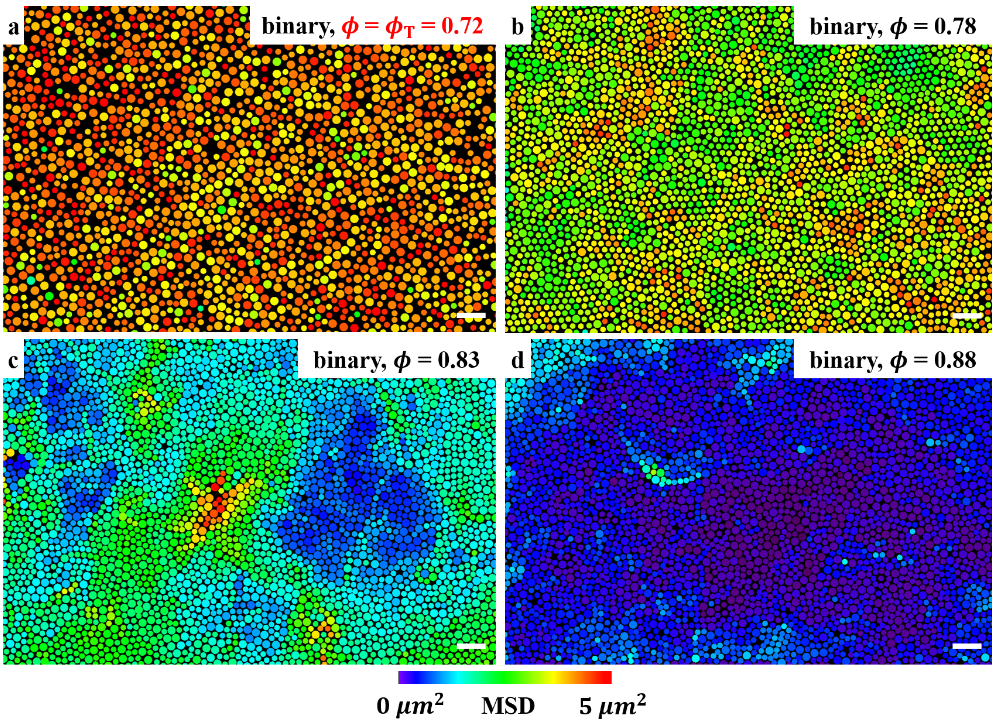}
	\caption{Dynamical heterogenity of binary spheres under \textbf{a} $\phi=\phi_\textrm{T}=0.72$, \textbf{b} $\phi=0.78$, \textbf{c} $\phi=0.83$, and \textbf{d} $\phi=0.88$. The color represents the value of the mean square displacement of a particle during 500~$s$. The scale bar is 10~$\mu$m.}
	\label{fig:S4}
\end{figure}

\renewcommand{\thefigure}{S5}
\begin{figure}[!h]
	\centering
	\includegraphics[width=1.0\columnwidth]{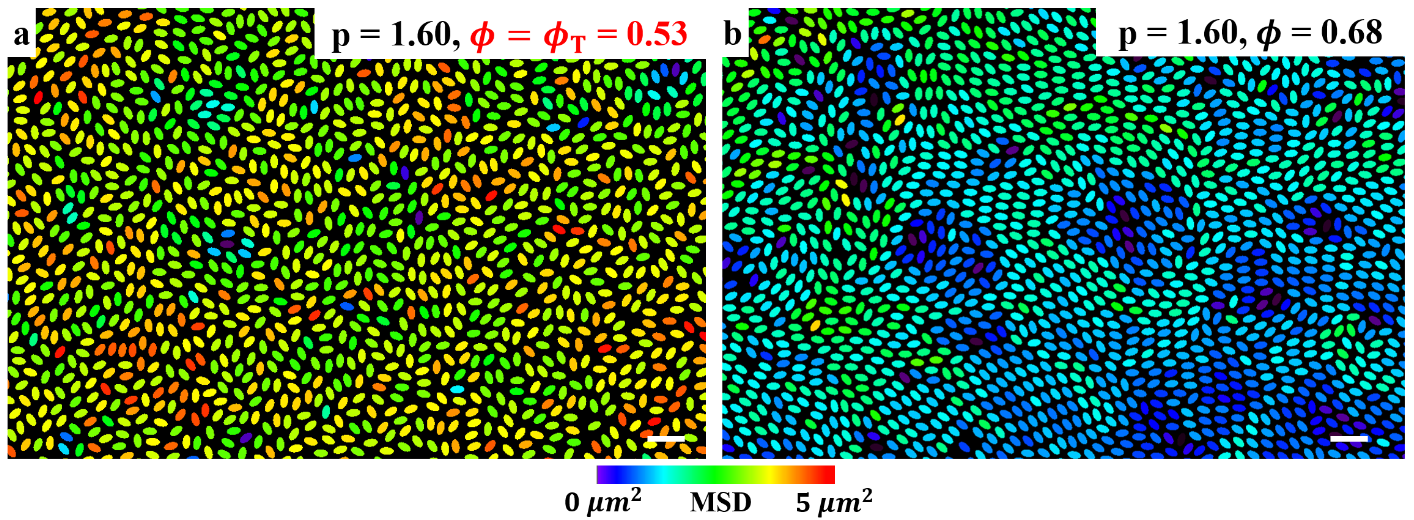}
	\caption{Dynamical heterogenity of $p=1.60$ ellipsoids under \textbf{a} $\phi=\phi_\textrm{T}=0.53$ and \textbf{b} $\phi=0.68$. The color represents the value of the mean square displacement of a particle during 500~$s$. The scale bar is 10~$\mu$m.}
	\label{fig:S5}
\end{figure}

In the main text of the article, we argue that the turning point of $D^{T}-s_\textrm{2}^{T}$ scaling was irrelevant to the dynamical heterogeneity (DH) by comparing the turning point ($\phi_\textrm{T}$) with the onset of DH. We found that for ellipsoids with $p=1.84, 2.68, 3.72, 5.00, 7.06$, $\phi_\textrm{T}$ was too small for any cluster to form within the system (Fig.~2). It can be simply judged from the raw images (Fig.~2d-h) that DH was not able to cause the turning in $D^{T}-s_\textrm{2}^{T}$ scaling in these systems. However, because $\phi_\textrm{T}$ values for binary and ellipsoids with $p=1.26, 1.60$ were relatively high ($\phi_\textrm{T}=0.72, 0.75$, and $0.53$, respectively), more analysis is required to rule out the role DH might have played in the turning of $D^{T}-s_\textrm{2}^{T}$ scaling in these three systems. Therefore, here, we discuss the relationship between DH and turning in $D^{T}-s_\textrm{2}^{T}$ scaling.

\renewcommand{\thefigure}{S6}
\begin{figure}[!h]
	\centering
	\includegraphics[width=1.0\columnwidth]{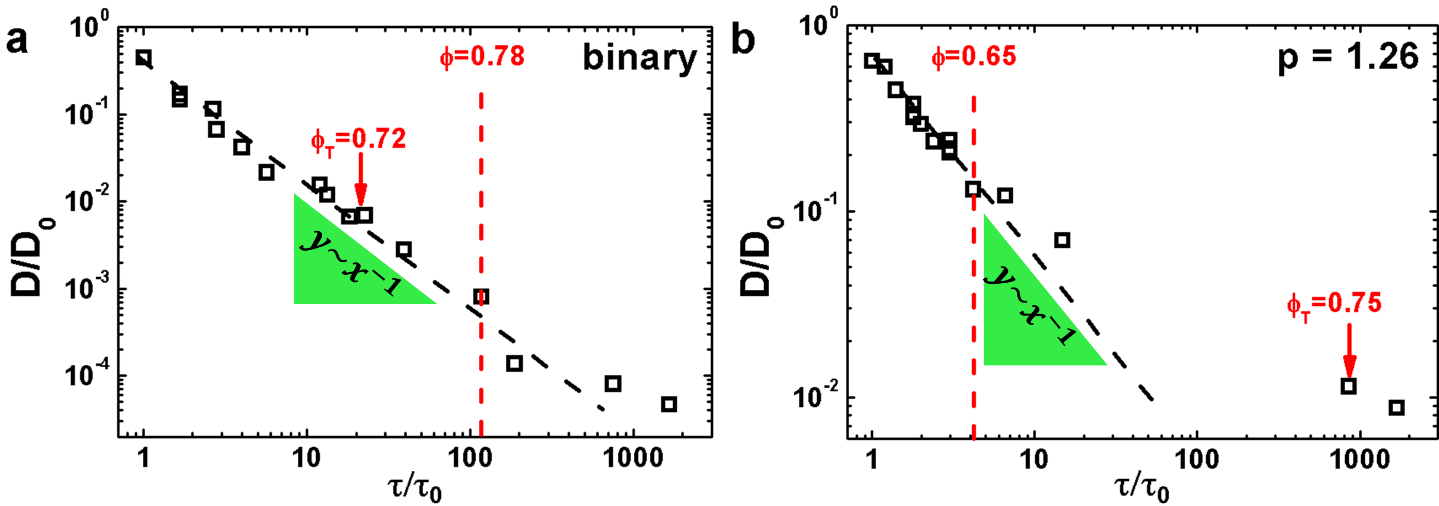}
	\caption{Rescaled self-diffusion coefficient $D/D_\textrm{0}$ as a function of rescaled relaxation time $\tau_{\alpha}/\tau_{\alpha}^{0}$ for \textbf{a} binary spheres and \textbf{b} $p=1.26$ ellipsoids. The dashed lines are fits of the data at low $\phi$ to the equation $y\sim x^{-1}$. The red arrows indicate the data point corresponding to the turning of $D^{T}-s_\textrm{2}^{T}$ scaling, $\phi_\textrm{T}$. The vertical dashed lines indicate the onset of DH.}
	\label{fig:S6}
\end{figure}

For $p=1.26$ ellipsoids, DH occurred before the turning of $D^{T}-s_\textrm{2}^{T}$ scaling. At $\phi=0.65$, particles with fast translational motion started to cluster and separated from slow particles in real space (Fig.~S3a). The length scale of the separated clusters grew observably with $\phi$ thereafter (Fig.~S3b-d). At $\phi_\textrm{T}=0.75$, when the $D^{T}-s_\textrm{2}^{T}$ scaling had broken down, the characteristics of DH were unambiguously observable throughout the whole system (Fig.~S3b). Therefore, the occurrence of DH cannot have been responsible for the turning of $D^{T}-s_\textrm{2}^{T}$ scaling in $p=1.26$ ellipsoids.

\renewcommand{\thefigure}{S7}
\begin{figure}[!h]
	\centering
	\includegraphics[width=1.0\columnwidth]{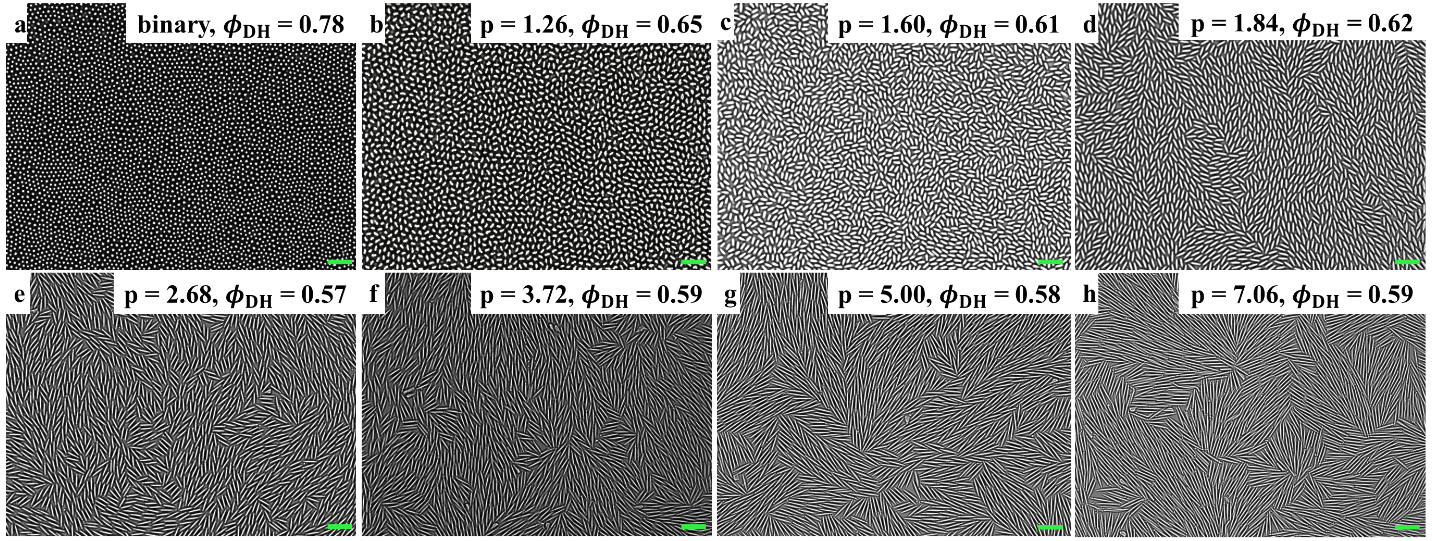}
	\caption{Raw images at the breakdown points of the Sokes$\textrm{-}$Einstein relation. The area fractions ($\phi_\textrm{DH}$) corresponding to the breakdown points are specified in each figure. The scale bar is 5~$\mu$m.}
	\label{fig:S7}
\end{figure}

For binary spheres and $p=1.60$ ellipsoids, DH occurred later than the turning of $D^{T}-s_\textrm{2}^{T}$ scaling. At $\phi_\textrm{T}$, the distribution of particle mobility was completely homogeneous for both systems (Fig.~S4a and Fig.~S5a). DH did not occur until the area fractions were more than $10\%$ higher than $\phi_\textrm{T}$ (Fig.~S4b-d and Fig.~S5b). Therefore, the DH played no role in the turning of $D^{T}-s_\textrm{2}^{T}$ scaling for either binary spheres or $p=1.60$ ellipsoids.

An immediate consequence of DH is the breakdown of the Stokes$\textrm{-}$Einstein (SE) relation $D^{T}=k_\textrm{B}T/6\pi \eta a$ \cite{2000JPCM_Angell}, which has been proven to be a hallmark of simple liquids \cite{2014PRL_Flenner,2015PRL_Mishra}. The superscript $T$ denotes the translational degree of freedom, $a$ is the particle radius, and $k_\textrm{B}T$ is the thermal energy. Because the viscosity $\eta$ is always proportional to the relaxation time $\tau_{\alpha}$ of the system, it is expected that $D^{T}$ is proportional to $\tau_{\alpha}^{-1}$ if the dynamics of the system is homogeneous. It was reported from both a simulation \cite{2014PRL_Flenner} and a colloidal experiment \cite{2015PRL_Mishra} that onset of DH was directly associated with the breakdown of the SE relation.

To clarify the relation between onset of DH and the $D^{T}-s_\textrm{2}^{T}$ turning point, we investigated the SE relation in our glass$\textrm{-}$forming liquids; the results of binary spheres and $p=1.26$ ellipsoids are presented in Fig.~S6. In accordance with the real space analysis (Fig.~S3 and Fig.~S4), we found that $\phi_\textrm{T}$ was smaller than the breakdown point of the SE relation for binary spheres (Fig.~S6a) and larger for $p=1.26$ ellipsoids (Fig.~S6b). Through the same method, we found that for the other six ellipsoid systems, each $\phi_\textrm{T}$ was far lower than the breakdown point of the SE relation, or equivalently, the onset of DH. For direct comparison with Fig.~2, the raw images corresponding to the breakdown point of the SE relation are presented in Fig.~S7. Unambiguously, $\phi_\textrm{T}$ and $\phi_\textrm{DH}$ differed substantially from each other.  

We further tested the spatial distribution of $s_\textrm{2}^{T}$ for binary spheres and $p=1.26$ and $p=1.60$ ellipsoids. The distributions became highly heterogeneous at high values of $\phi$ (Fig.~S8); and in a manner similar to the case of DH, the $s_\textrm{2}^{T}$ heterogeneity (SH) was irrelevant to the $D^{T}-s_\textrm{2}^{T}$ turning because the distribution of $s_\textrm{2}^{T}$ was still homogeneous at $\phi_\textrm{T}$ (Fig.~S8a,c,e). In accordance with previous observations \cite{2014NC_Zheng, 2010NM_Tanaka, 2016NP_Golde}, although  the correlation between structural heterogeneity and DH existed over $\phi$, their spatial correlation was weak in our experiments.

The analysis regarding DH and SH further strengthens the argument in the main text that DH is irrelevant to the turning of $D^{T}-s_\textrm{2}^{T}$ scaling. Majority of glass$\textrm{-}$forming liquids remained completely homogeneous in both dynamics and structure at $\phi_\textrm{T}$. For the exception, $p=1.26$ ellipsoids, both DH and SH were established before $\phi_\textrm{T}$. As discussed in the main text, the slowing down of kinetics, in our observation, is directly controlled by the inherent structure of the system. A deep understanding of the turning thus requires a detailed description of the $\Phi$-scape.

\newpage
\renewcommand{\thefigure}{S8}
\begin{figure}[!h]
	\centering
	\includegraphics[width=1.0\columnwidth]{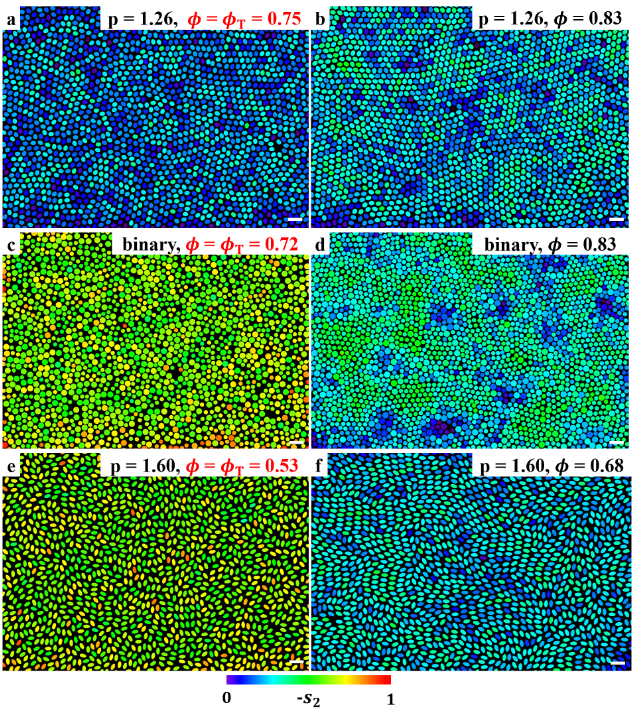}
	\caption{SH of \textbf{a, b} $p=1.26$ ellipsoids, \textbf{c, d} binary spheres, and \textbf{e, f} $p=1.60$ ellipsoids. The color represents the value of the excess entropy of a particle. The scale bar is 10~$\mu$m.}
	\label{fig:S8}
\end{figure}

\section{$\Phi$-scape topography}

\renewcommand{\thefigure}{S9}
\begin{figure}[!h]
	\centering
	\includegraphics[width=0.45\columnwidth]{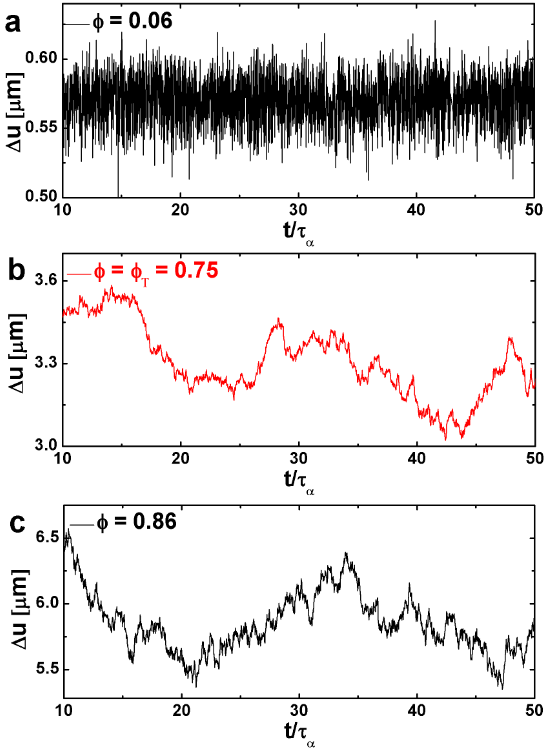}
	\caption{Topography of the $\Phi$-scape in representation of particle mobility for $p=1.26$ ellipsoid system. We chose \textbf{a}, $\phi=0.06$, \textbf{b}, $\phi=\phi_\textrm{T}=0.75$, and \textbf{c}, $\phi=0.86$ samples and calculated the time evolution of their displacement, $\Delta u(t)$, during the relaxation time $\tau_{\alpha}$. $\tau_{\alpha}$ was defined as the time at which the dynamical structure factor in Fig.~5a had decayed to $1/e$. The mobility in rotational motion is missing because, for $p=1.26$ ellipsoid, the determination of particles' orientation was inaccurate when they slightly deviated from xy-plane.} 
	\label{fig:S9}
\end{figure}

\renewcommand{\thefigure}{S10}
\begin{figure}[!h]
	\centering
	\includegraphics[width=1.0\columnwidth]{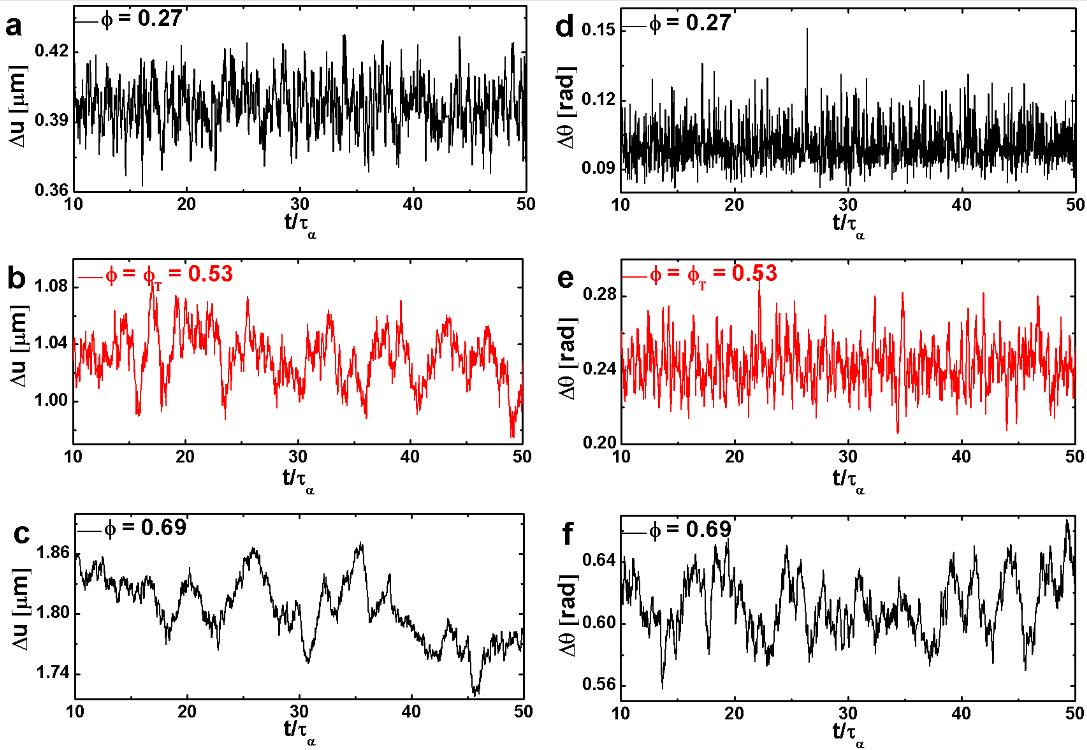}
	\caption{Topography of the $\Phi$-scape in representation of particle mobility for $p=1.60$ ellipsoid system. We chose \textbf{a}, $\phi=0.27$, \textbf{b}, $\phi=\phi_\textrm{T}=0.53$, and \textbf{c}, $\phi=0.69$ samples and calculated the time evolution of their displacement, $\Delta u(t)$, during the relaxation time $\tau_{\alpha}$. \textbf{d-f,} The rotational mobility $\Delta \theta (t)$ under the same $\phi$. $\tau_{\alpha}$ was defined as the time at which the dynamical structure factor in Fig.~5a had decayed to $1/e$.} 
	\label{fig:S10}
\end{figure}

\renewcommand{\thefigure}{S11}
\begin{figure}[H]
	\centering
	\includegraphics[width=1.0\columnwidth]{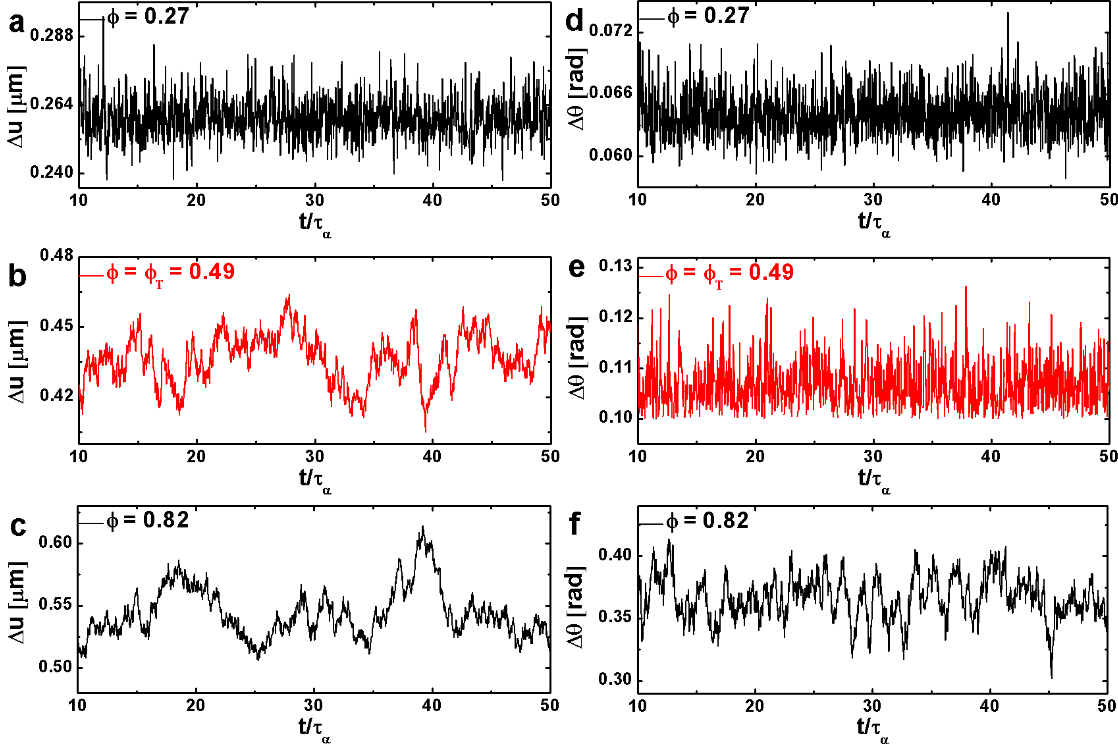}
	\caption{Topography of the $\Phi$-scape in representation of particle mobility for $p=1.84$ ellipsoid system. We chose \textbf{a}, $\phi=0.27$, \textbf{b}, $\phi=\phi_\textrm{T}=0.49$, and \textbf{c}, $\phi=0.82$ samples and calculated the time evolution of their displacement, $\Delta u(t)$, during the relaxation time $\tau_{\alpha}$. \textbf{d-f,} The rotational mobility $\Delta \theta (t)$ under same $\phi$. $\tau_{\alpha}$ was defined as the time when the dynamical structure factor in Fig.~5a had decayed to $1/e$.} 
	\label{fig:S11}
\end{figure}

\renewcommand{\thefigure}{S12}
\begin{figure}[H]
	\centering
	\includegraphics[width=1.0\columnwidth]{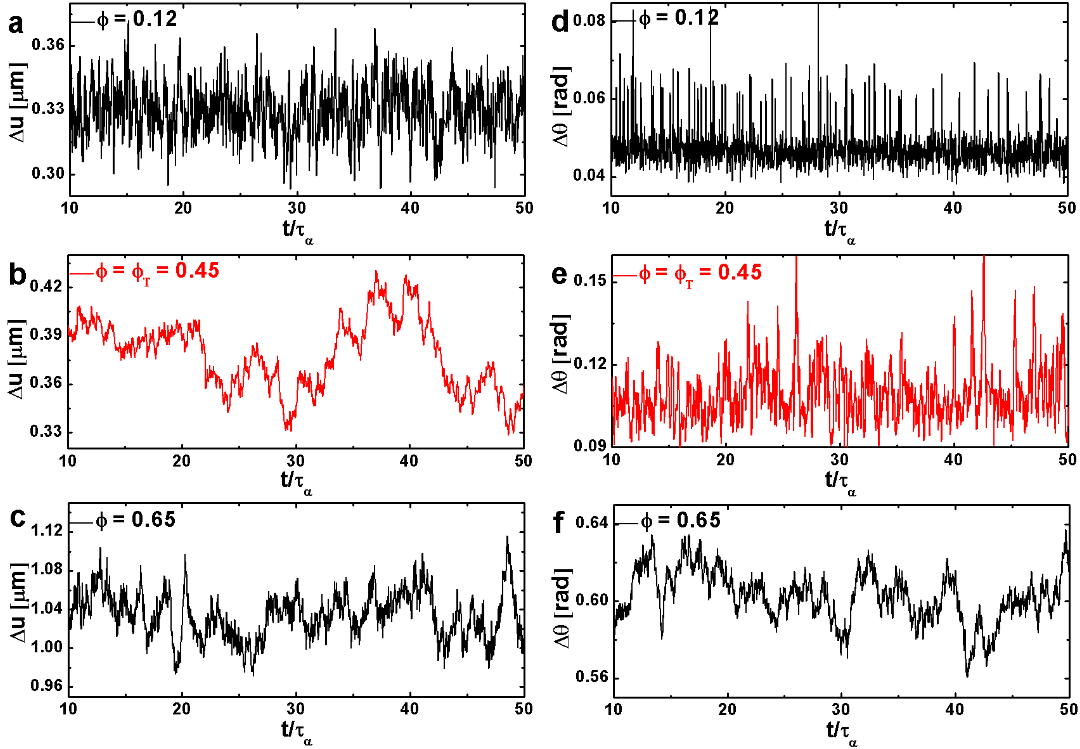}
	\caption{Topography of the $\Phi$-scape in representation of particle mobility for $p=2.68$ ellipsoid system. We chose \textbf{a}, $\phi=0.12$, \textbf{b}, $\phi=\phi_\textrm{T}=0.45$, and \textbf{c}, $\phi=0.65$ samples and calculated the time evolution of their displacement, $\Delta u(t)$, during the relaxation time $\tau_{\alpha}$. \textbf{d-f,} The rotational mobility $\Delta \theta (t)$ under same $\phi$. $\tau_{\alpha}$ was defined as the time when the dynamical structure factor in Fig.~5a had decayed to $1/e$.} 
	\label{fig:S12}
\end{figure}

\renewcommand{\thefigure}{S13}
\begin{figure}[H]
	\centering
	\includegraphics[width=1.0\columnwidth]{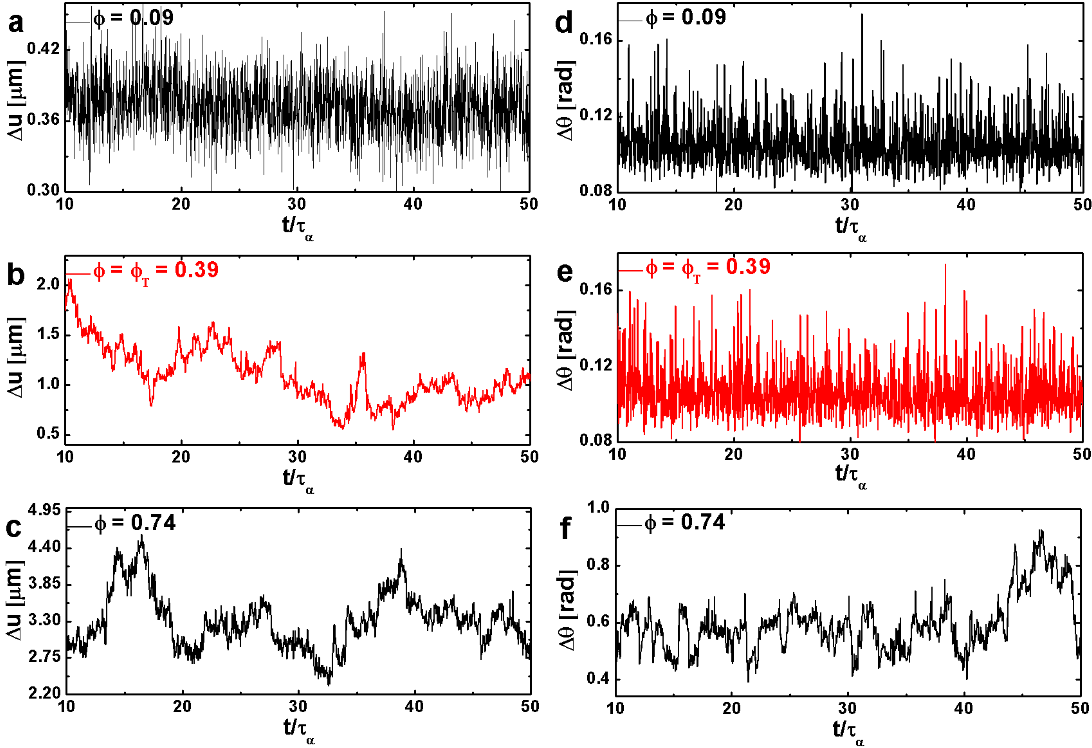}
	\caption{Topography of the $\Phi$-scape in representation of particle mobility for $p=3.72$ ellipsoid system. We chose \textbf{a}, $\phi=0.09$, \textbf{b}, $\phi=\phi_\textrm{T}=0.39$, and \textbf{c}, $\phi=0.74$ samples and calculated the time evolution of their displacement, $\Delta u(t)$, during the relaxation time $\tau_{\alpha}$. \textbf{d-f,} The rotational mobility $\Delta \theta (t)$ under same $\phi$. $\tau_{\alpha}$ was defined as the time when the dynamical structure factor in Fig.~5a had decayed to $1/e$.} 
	\label{fig:S13}
\end{figure}

\renewcommand{\thefigure}{S14}
\begin{figure}[H]
	\centering
	\includegraphics[width=1.0\columnwidth]{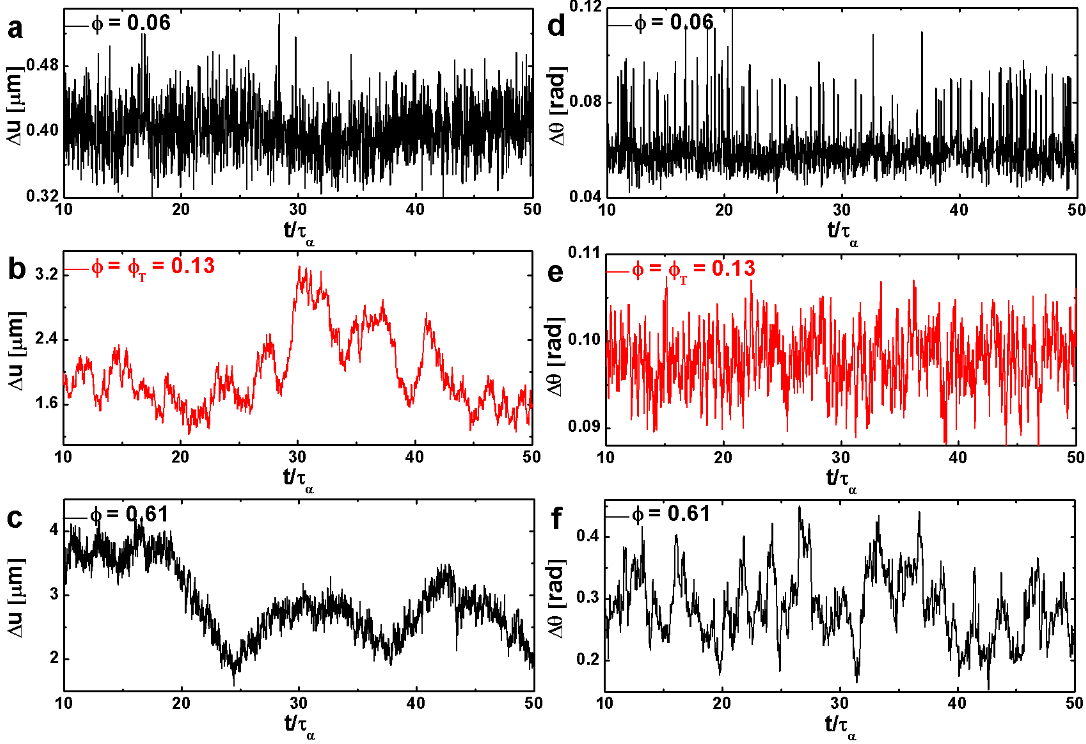}
	\caption{Topography of the $\Phi$-scape in representation of particle mobility for $p=7.06$ ellipsoid system. We chose \textbf{a}, $\phi=0.06$, \textbf{b}, $\phi=\phi_\textrm{T}=0.13$, and \textbf{c}, $\phi=0.61$ samples and calculated the time evolution of their displacement, $\Delta u(t)$, during the relaxation time $\tau_{\alpha}$. \textbf{d-f,} The rotational mobility $\Delta \theta (t)$ under same $\phi$. $\tau_{\alpha}$ was defined as the time when the dynamical structure factor in Fig.~5a had decayed to $1/e$.} 
	\label{fig:S14}
\end{figure}

In the main text, we report that the topography of the $\Phi$-scape in translational motion started to become rugged in the vicinity of $\phi_\textrm{T}$ for both $p=5.00$ ellipsoids (Fig.~4a-c) and binary systems (Fig.~4g-i) whereas it remained smooth in rotational motion (Fig.~4d-f) for $p=5.00$ ellipsoids. Based on this observation, it was concluded that the turning of $D^{T}-s_\textrm{2}^{T}$ scaling was caused by a change in the $\Phi$-scape topography for translational motion. Here, we provide the translational and rotational $\Phi$-scape topographies in mobility representation for all other ellipsoids (Figs.~S9$\textrm{-}$S14). Without exception, the $D^{T}-s_\textrm{2}^{T}$ plot turned when the $\Phi$-scape became rugged in translational motion. For rotational motion, no larger$\textrm{-}$scaled structures or basins were observed in the $\Phi$-scape until area fractions were close to $\phi_\textrm{g}$, which was consistent with the absence of turnings in the $D^{\theta}-s_\textrm{2}^{\theta}$ plots.

\section{Structural features of strong and fragile liquids}

\renewcommand{\thefigure}{S15}
\begin{figure}[!h]
	\centering
	\includegraphics[width=1.0\columnwidth]{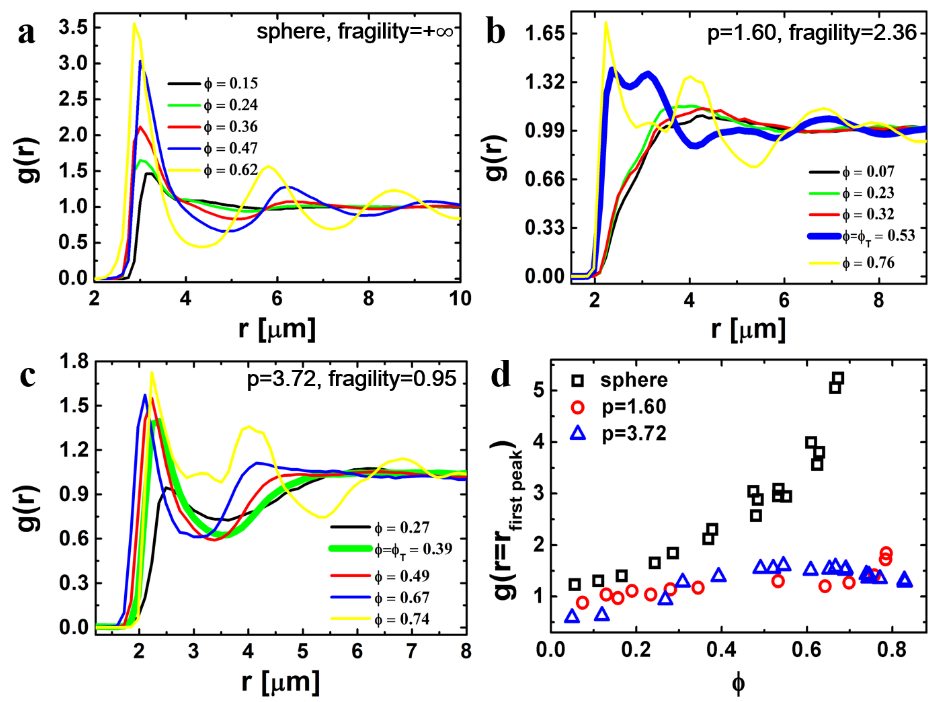}
	\caption{Radial distribution function $g(r)$ for \textbf{a} sphere, \textbf{b} $p=1.60$ ellipsoid, and \textbf{c} $p=3.72$ ellipsoid systems. \textbf{d}, first peak value of $g(r)$ as a function of $\phi$. The bold curves in \textbf{b, c} represent g(r) at $\phi_\textrm{T}$.}
	\label{fig:S15}
\end{figure}

The distinction between strong and fragile liquids can be vividly appreciated from their structures, because liquid structure reacts to cooling in a manner that is dependent on the liquid's fragility. For a strong liquid, cooling can not alter the structure considerably \cite{2014NC_Mauro}. As a result, the radial distribution functions of a strong liquid at high $\phi$ remained similar to that at low $\phi$ (Fig.~S15c); this differed from the case of fragile liquids (Fig.~S15a, b). The evolution of the first $g(r)$ peak value (Fig.~S15d) quantitatively demonstrates this difference. For the strong liquid, containing $p=3.72$ ellipsoids, the fragility of which was smaller than 1, the value of $g(r=r_\textrm{first peak})$ even slightly decreased with $\phi$ in the high$\textrm{-}$density region. For spherical system, $g(r=r_\textrm{first peak})$ increased dramatically as $\phi$ approached $\phi_\textrm{m}$. Note that after the melting point, $\phi_\textrm{m}$($<\phi_\textrm{g}$), the system crystalised and the relaxation time jumped to infinity. Therefore, we can conclude that the structure of a c-path liquid is much more sensitive to the cooling process than that of a g-path liquid.

\end{document}